\documentstyle[12pt]{article}

\catcode`\@=11
\long\def\@makefntext#1{
\protect\noindent \hbox to 3.2pt {\hskip-.9pt
$^{{\ninerm\@thefnmark}}$\hfil}#1\hfill}                

\def\@makefnmark{\hbox to 0pt{$^{\@thefnmark}$\hss}}  

\def\ps@myheadings{\let\@mkboth\@gobbletwo
\def\@oddhead{\hbox{}
\rightmark\hfil\ninerm\thepage}
\def\@oddfoot{}\def\@evenhead{\ninerm\thepage\hfil
\leftmark\hbox{}}\def\@evenfoot{}
\def\sectionmark##1{}\def\subsectionmark##1{}}

\renewcommand{\thefootnote}{\fnsymbol{footnote}}

\def\sectionc{\@startsection {section}{1}{\z@}{-3.5ex plus -1ex minus 
    -.2ex}{2.3ex plus .2ex}{\bf }}
\def\subsectionc{\@startsection{subsection}{2}{\z@}{-3.25ex plus -1ex minus 
   -.2ex}{1.5ex plus .2ex}{\it }}
\renewcommand{\section}[1]{\sectionc{#1}\hspace*{\parindent}}
\renewcommand{\subsection}[1]{\subsectionc{#1}\hspace*{\parindent}}
\newcounter{appendixc}
\newcounter{subappendixc}[appendixc]
\newcounter{subsubappendixc}[subappendixc]

\renewcommand{\appendix}[1] {\vspace*{0.6cm}
        \refstepcounter{appendixc}
        \setcounter{figure}{0}
        \setcounter{table}{0}
        \setcounter{equation}{0}
        \renewcommand{\thefigure}{\Alph{appendixc}.\arabic{figure}}
        \renewcommand{\thetable}{\Alph{appendixc}.\arabic{table}}
        \renewcommand{\theappendixc}{\Alph{appendixc}}
        \renewcommand{\theequation}{\Alph{appendixc}.\arabic{equation}}
        \noindent{\bf Appendix \theappendixc #1}\par\vspace*{0.4cm}}

\def\abstracts#1{{
        \centering{\begin{minipage}{13.2truecm}\footnotesize\baselineskip=13pt\noindent
        \parindent=0pt #1
        \end{minipage}}\par}}

\renewenvironment{thebibliography}[1]
        {\begin{list}{\arabic{enumi}.}
        {\usecounter{enumi}\setlength{\parsep}{0pt}
\setlength{\leftmargin 0.75cm}{\rightmargin 0pt}
         \setlength{\itemsep}{0pt} \settowidth
        {\labelwidth}{#1.}\sloppy}}{\end{list}}

\newcounter{itemlistc}
\newcounter{romanlistc}
\newcounter{alphlistc}
\newcounter{arabiclistc}

\newcommand{\fcaption}[1]{
        \refstepcounter{figure}
        \setbox\@tempboxa = \hbox{\footnotesize Figure~\thefigure. #1}
        \ifdim \wd\@tempboxa > 6in
           {\begin{center}
        \parbox{6in}{\footnotesize\baselineskip=13pt Figure~\thefigure. #1}
            \end{center}}
        \else
             {\begin{center}
             {\footnotesize Figure~\thefigure. #1}
              \end{center}}
        \fi}

\newcommand{\tcaption}[1]{
        \refstepcounter{table}
        \setbox\@tempboxa = \hbox{\footnotesize Table~\thetable. #1}
        \ifdim \wd\@tempboxa > 6in
           {\begin{center}
        \parbox{6in}{\footnotesize\baselineskip=13pt Table~\thetable. #1}
            \end{center}}
        \else
             {\begin{center}
             {\footnotesize Table~\thetable. #1}
              \end{center}}
        \fi}

\def\@citex[#1]#2{\if@filesw\immediate\write\@auxout
        {\string\citation{#2}}\fi
\def\@citea{}\@cite{\@for\@citeb:=#2\do
        {\@citea\def\@citea{,}\@ifundefined
        {b@\@citeb}{{\bf ?}\@warning
        {Citation `\@citeb' on page \thepage \space undefined}}
        {\csname b@\@citeb\endcsname}}}{#1}}

\newif\if@cghi
\def\cite{\@cghitrue\@ifnextchar [{\@tempswatrue
        \@citex}{\@tempswafalse\@citex[]}}
\def\citelow{\@cghifalse\@ifnextchar [{\@tempswatrue
        \@citex}{\@tempswafalse\@citex[]}}
\def\@cite#1#2{{$\null^{#1}$\if@tempswa\typeout
        {IJCGA warning: optional citation argument
        ignored: `#2'} \fi}}

 1
 1
 1

\font\ninerm=cmr9

\begin{document}

\centerline{\normalsize\bf THE NUCLEAR-MATTER RESPONSE IN THE}
\baselineskip=15pt
\centerline{\normalsize\bf QUARK STRING-FLIP MODEL}

\vspace*{0.6cm}
\centerline{\footnotesize J. PIEKAREWICZ}
\baselineskip=13pt
\centerline{\footnotesize\it Supercomputer Computations Research Institute}
\baselineskip=13pt
\centerline{\footnotesize\it Florida State University, 
			     Tallahassee, FL 32306, USA}
\baselineskip=13pt
\centerline{\footnotesize E-mail: jorgep@scri.fsu.edu}

\vspace*{0.6cm}
\abstracts{Nuclear matter is modeled directly in terms of its
constituent quarks. A many-body string-flip potential is used that
confines quarks within hadrons, enables the hadrons to separate
without generating van der Waals forces, and is symmetric in all quark
coordinates. We present variational Monte Carlo results for the
ground-state properties of large, three-dimensional systems. A phase
transition from nuclear to quark matter is observed which is
characterized by a dramatic rearrangement of strings. We report on
exact calculations of the dynamic response of many-quark systems in
one spatial dimension. At low density and small momentum transfers the
response is substantially larger than that of a free Fermi gas of
quarks; this suggests that there is a coherent response from all the
quarks inside the hadron. This coherence, however, is incomplete, as
the response is suppressed relative to that of a free Fermi gas of
nucleons due to the internal quark substructure of the hadron.}

\normalsize\baselineskip=15pt
\setcounter{footnote}{0}
\renewcommand{\thefootnote}{\alph{footnote}}

\section{Introduction}\label{sec:intro}
 The advent of new, high-energy facilities for nuclear physics
research offers a unique possibility at identifying departures 
from a conventional, meson-baryon description of nuclear phenomena.
Naively, one would expect that as the typical distance scales probed
become short relative to the nucleon size, quarks and gluons should
become manifest in nuclear observables. Yet, quark-gluon signatures in
nuclear phenomena have, so far, proved elusive.  One of the objections
most often raised by the critics of conventional nuclear structure
models is that, because the intrinsic size of the hadrons, a picture
of nucleons interacting in the medium via meson exchanges is
inappropriate. Yet, there seems to be ample experimental evidence that
will support that, although some properties of the nucleon may be
modified in the medium, a nucleon inside the medium resembles to a
very good approximation a nucleon in free space. Perhaps one of the
greatest challenges facing nuclear physics today is the explanation of
these remarkable facts: why are these effective hadronic models so 
successful and how can such models emerge from the basic underlying
theory having quarks and gluons as the fundamental degrees of freedom.
It is also interesting to note that, although most of these questions
have been posed since the advent of QCD, little progress has been made
in answering them. A serious difficulty encountered in attempting to
answer these questions is how to model a system that is believed to
have quarks confined inside color-neutral hadrons at low density but
free quarks at high density. The divorce of the two pictures is
perhaps forced by the difficulty of treating quark confinement:
how can quarks be confined inside hadrons, yet hadrons can 
separate without generating long-range van der Waals forces? 
We offer no new insights into this difficult problem; rather
we argue that it is useful to consider an effective model which 
interpolates between a hadron- and quark-based description at 
low and high density.

 In this contribution we consider nuclear matter from the viewpoint 
of the constituent quark model. The motivation for this study is
threefold. First, we wish to examine the role of nucleon substructure 
in nuclear observables. For example, we wish to understand how are 
hadronic properties---such as the nucleon form factor---modified in 
the nuclear environment. Further, we wish to understand, as a function
of the density and the momentum transfer, when do leptons scatter from
nucleons and when do they scatter from individual quarks. Second, we 
want to identify signatures for the nuclear- to quark-matter
transition. For example, how is the color susceptibility modified as 
a function of the density of the system. Finally, we want to search
for qualitatively new modes of excitation in many-quark systems. These
modes---which could be collective excitations of several quarks in the
nucleus---are not present in single hadrons nor in hadronic models of 
the nucleus. These ``quark giant resonances'' could involve the
coherent response of many quarks to density, spin, flavor, or color 
fluctuations. Moreover, we are interested in studying the mixing 
between quark-like excitations, such as the $N \rightarrow \Delta$ 
transition, and hadronic excitations, such as the Gamow-Teller
resonance. Indeed, it is unknown how the excitation energy and mixing 
of these modes will change with density.

\section{The String-flip Model}\label{sec:sfmodel}
 The string-flip, or quark-exchange, model\cite{lenz86} is a simple 
many-body generalization of the nonrelativistic constituent quark 
model\cite{isgkar79}. Yet the obvious generalization consisting of 
pairwise confining forces between quarks is known to generate
long-range van der Waals forces\cite{grelip81}. These long-range 
forces which are power-law---rather than exponentially---suppressed 
at long distances do not exist in nature. Rather, it is cluster
separability, namely, the possibility for color-singlet hadrons to 
separate without generating residual van der Waals forces, that is 
observed in nature. The string-flip model succeeds in providing
cluster separability due to the intrinsic many-body nature of the 
potential\cite{hmn85,watson89,horpie91,horpie92,fripie94,alber92}. 

\begin{figure}[h]
\begin{center}
 \null
 \vskip1.2in
 \includegraphics{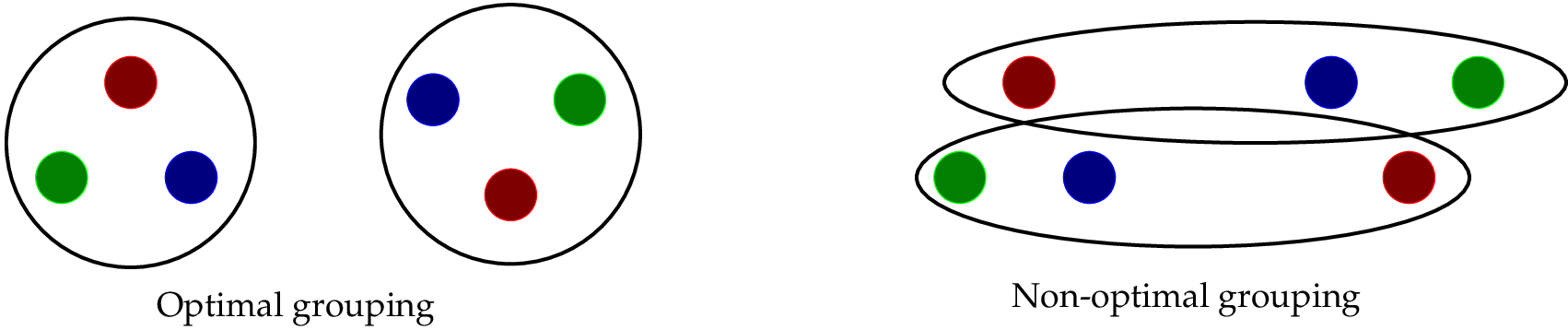}
\fcaption{Optimal and non-optimal grouping of quarks into hadrons.}
\label{figone}
\end{center}
\end{figure}

\begin{figure}[h]
\begin{center}
 \null
 \vskip1.2in
 \includegraphics{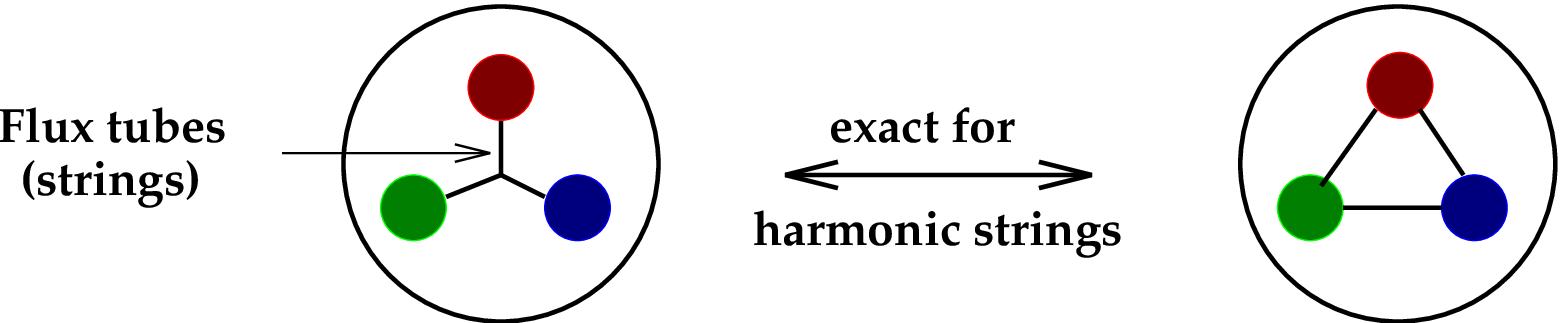}
\fcaption{Adiabatic approximation to the complicated quark-gluon
	  dynamics.}
\label{figtwo}
\end{center}
\end{figure}

 The Hamiltonian for the present version of the string-flip model 
is given by\cite{horpie92,fripie94}
\begin{equation}
   H = \sum_{i=1}^{N} {P_{i}^{2} \over 2m} + 
   V({\bf r}_{\lower 2pt \hbox{$\scriptstyle 1$}},
   \ldots,{\bf r}_{\lower 2pt \hbox{$\scriptstyle N$}}) \;.
 \label{hamil}
\end{equation}
We consider equal numbers ($A=N/3$) of red, blue, and green quarks; 
for simplicity we assume that quarks are devoid of any additional 
(spin and flavor) intrinsic degree of freedom. We require each
color-singlet hadron to be formed by one red, one blue, and one
green quark. Moreover, we require the grouping of quarks into 
color-singlet hadrons to be optimal. That is (we use units
in which $k=m=\hbar=1$)
\begin{equation}
 V({\bf r}_{\lower 2pt \hbox{$\scriptstyle 1$}},
 \ldots,{\bf r}_{\lower 2pt \hbox{$\scriptstyle N$}}) =
 {\rm Min}
 \sum_{i=1}^{A} {1 \over 2} \Big[
 ({\bf r}^{\scriptscriptstyle R}_{\lower 2pt \hbox{$\scriptstyle i$}} -
  {\bf r}^{\scriptscriptstyle B}_{\lower 2pt \hbox{$\scriptstyle i$}})^{2} +
 ({\bf r}^{\scriptscriptstyle B}_{\lower 2pt \hbox{$\scriptstyle i$}} -
  {\bf r}^{\scriptscriptstyle G}_{\lower 2pt \hbox{$\scriptstyle i$}})^{2} +
 ({\bf r}^{\scriptscriptstyle G}_{\lower 2pt \hbox{$\scriptstyle i$}} -
  {\bf r}^{\scriptscriptstyle R}_{\lower 2pt \hbox{$\scriptstyle i$}})^{2} 
  \Big] \;.
 \label{vtqc}
\end{equation}
Here ${\bf r}^{\scriptscriptstyle R}_{\lower 2pt \hbox{$\scriptstyle i$}}$,
${\bf r}^{\scriptscriptstyle B}_{\lower 2pt \hbox{$\scriptstyle i$}}$, and
${\bf r}^{\scriptscriptstyle G}_{\lower 2pt \hbox{$\scriptstyle i$}}$,
are the positions of the red, blue, and green quarks ``belonging'' to
the $i^{\rm th}$ hadron, and the minimization procedure is taken over 
a---potentially enormous---set of $(A!)^{2}$ possible groupings.
An example of an optimal and non-optimal grouping of six quarks into
two color-singlet hadrons is depicted in Fig.~\ref{figone}.
The many-body potential represents an adiabatic approximation to the 
extremely complicated quark-gluon dynamics. We assume that the gluons 
are ``light'' degrees of freedom with a dynamical time-scale that is 
much shorter than that of the ``heavy'' quarks; as quarks move,
the gluonic strings adjust instantaneously to the position of the
quarks. The basic problem, then, is to decide how quarks are to be 
grouped inside the hadrons: a gluonic string (or flux tube) leaving 
a quark must end up in a three-quark junction (see Fig.~\ref{figtwo}). 
The fundamental question becomes which junction? Presumably, lattice 
QCD solves this problem---but at a spectacular computational cost. It 
is because of these enormous computational demands that one must
resort to simple phenomenological models. Nevertheless, most models 
of the many-quark dynamics will need to determine which quarks belong 
to a particular hadron. Thus, solving some kind of
``quark-assignment'' problem is likely to be a general requirement 
for these models.
The potential is a many-body operator; it can not be reduced to a
sum of two-body terms. Indeed, the movement of a single quark might
cause several string to flip. Moreover, the many-body potential is 
symmetric under the exchange of all the quark coordinates---even 
those ``belonging'' to different hadrons. Quarks are also confined 
inside color-singlet hadrons. However, there are now no long-range 
van der Waals forces between the hadrons, as the force now saturates 
within the hadron. Thus, the only residual interaction between hadrons
involves the possibility of quark exchange and the Pauli principle 
between identical quarks. This immediately suggests that a typical 
distance scale characterizing the interaction is determined by the 
size of the hadron; if the distance between hadrons is much larger 
than this size, quark-exchange should be suppressed. Any long-range 
component to the hadron-hadron potential, arising phenomenologically 
from the exchange of pseudo-Goldstone bosons, is beyond the scope of 
this simple model. 

\section{Static Properties}\label{sec:static}
 We compute the ground-state properties of the many-quark system using
a variational Monte Carlo approach. The dynamics of the system is
contained in a simple one-parameter variational wave function of the
form\cite{horpie92,fripie94}:
\begin{equation}
  \Phi_{\lower 2pt \hbox{$\scriptstyle \lambda$}}
 ({\bf r}_{\lower 2pt \hbox{$\scriptstyle 1$}},\ldots,
  {\bf r}_{\lower 2pt \hbox{$\scriptstyle N$}}) =
  \exp\big[-\lambda V
 ({\bf r}_{\lower 2pt \hbox{$\scriptstyle 1$}},\ldots,
  {\bf r}_{\lower 2pt \hbox{$\scriptstyle N$}})\big]
  \Phi_{\rm FG}
 ({\bf r}_{\lower 2pt \hbox{$\scriptstyle 1$}},\ldots,
  {\bf r}_{\lower 2pt \hbox{$\scriptstyle N$}}) \;.
 \label{varwfn}
\end{equation}
The Fermi-gas wave function ($\Phi_{\rm FG}$) is a product of red, 
blue, and green Slater determinants. The Fermi-gas wave function is 
exact for a system of identical fermions with no correlations other 
than those generated by the Pauli exclusion principle. The exponential
factor, on the other hand, characterizes the amount of clustering in
the ground state through the variational parameter $\lambda$. For a
dilute system of quarks---where quarks cluster into individual 
color-singlet hadrons---the variational wave function reproduces the
exact wave function of isolated clusters in the limit of
$\lambda\rightarrow 1/\sqrt{3}$. Moreover, in the high-density 
limit---where the interparticle separation is substantially smaller
than the confinement scale---the potential energy becomes unimportant
and the variational wave function reproduces the exact Fermi-gas 
result in the limit of $\lambda\rightarrow 0$. Thus, this simple
one-parameter variational wave function is exact in the low- and
high-density limits, with $\lambda^{-1/2}$ playing the role of a
confinement scale. One of the advantages of using such a simple
variational wave function is that the expectation value of the
kinetic and potential energies are not independent:
\begin{equation}
  \langle \Phi_{\lambda} |T|\Phi_{\lambda} \rangle = 
  T_{\rm FG} + 3\lambda^{2}
  \langle \Phi_{\lambda} |V|\Phi_{\lambda} \rangle \;.
 \label{ekinet}
\end{equation}
Here $T_{\rm FG}$ is the kinetic energy of a free Fermi gas and
the additional term, $3\lambda^{2}\langle V \rangle$, represents 
the increase in kinetic energy above the Fermi-gas limit due to 
the presence of clustering correlations. Thus, to compute the total 
energy of the system we only need to evaluate  the expectation value 
of the potential energy 
\begin{equation}
  E_{\lambda}(\rho) \equiv
  \langle \Phi_{\lambda} |H| \Phi_{\lambda} \rangle = 
  T_{\rm FG} + (3\lambda^{2} +1)
  \langle \Phi_{\lambda} |V| \Phi_{\lambda} \rangle \;.
 \label{etotal}
\end{equation}
This expectation value is relatively simple to compute using
Metropolis Monte Carlo methods. Yet, one must realize that it
involves performing a $3N\simeq 300$-dimensional integral!

\begin{figure}[h]
\begin{center}
 \null
 \vskip1.1in
 \includegraphics{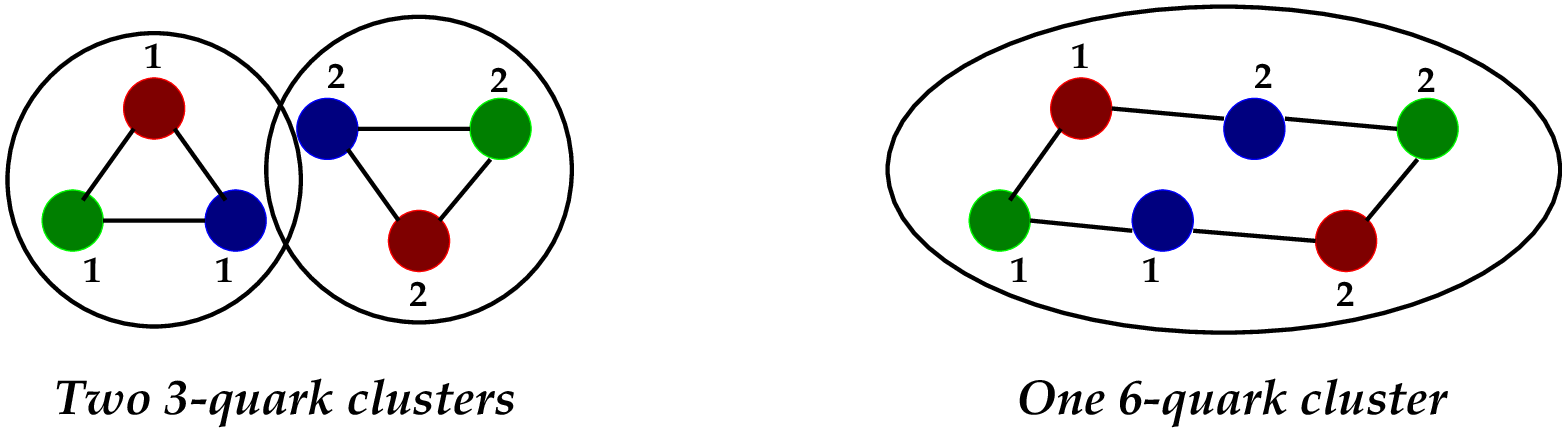}
\fcaption{Differences between the models due to possible 
          multi-quark configurations.}
\label{figthree}
\end{center}
\end{figure}

\begin{figure}[h]
\begin{center}
 \null
 \vskip1.0in
 \includegraphics{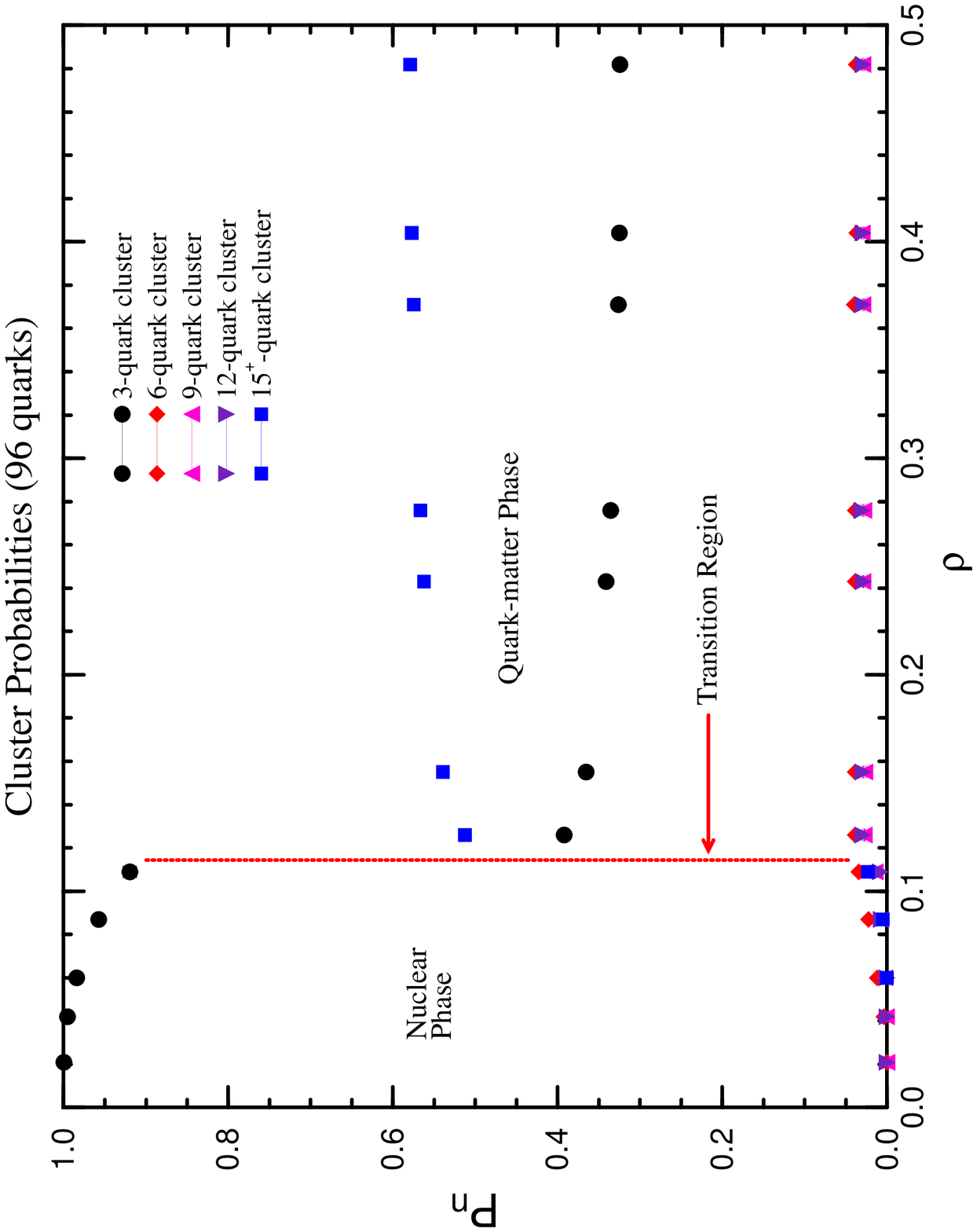}
 \vskip2.0in
\fcaption{Quark-cluster probabilities as a function of density.}
\label{figcluster}
\end{center}
\end{figure}

 The most demanding component of the calculation is the determination
of the optimal grouping of quarks into color-singlet hadrons. Indeed,
for the potential of Eq.~\ref{vtqc} there is no efficient (i.e.,
power-law) algorithm to solve this complicated assignment problem. 
Thus, we have resorted to the stochastic optimization technique of
simulated annealing\cite{fripie94}. We have computed the energy per 
quark and length-scale for quark confinement as a function of
density. Although these results already show how quark clustering 
decreases with density and characterize the nuclear- to quark-matter 
transition, they are limited to a very small ($A=8$) number of
hadrons. Thus, we find it advantageous to study a slightly different 
model but one in which the assignment problem can be solved readily. 
For this problem we consider pairing two colors at a time\cite{horpie92}. 
That is,
\begin{equation}
 V({\bf r}_{\lower 2pt \hbox{$\scriptstyle 1$}},
 \ldots,{\bf r}_{\lower 2pt \hbox{$\scriptstyle N$}}) =
 {\rm Min}
 \sum_{i=1}^{A} {1 \over 2} 
 ({\bf r}^{\scriptscriptstyle R}_{\lower 2pt \hbox{$\scriptstyle i$}} -
  {\bf r}^{\scriptscriptstyle B}_{\lower 2pt \hbox{$\scriptstyle i$}})^{2} +
 {\rm Min}
 \sum_{i=1}^{A} {1 \over 2} 
 ({\bf r}^{\scriptscriptstyle B}_{\lower 2pt \hbox{$\scriptstyle i$}} -
  {\bf r}^{\scriptscriptstyle G}_{\lower 2pt \hbox{$\scriptstyle i$}})^{2} +
 {\rm Min}
 \sum_{i=1}^{A} {1 \over 2} 
 ({\bf r}^{\scriptscriptstyle G}_{\lower 2pt \hbox{$\scriptstyle i$}} -
  {\bf r}^{\scriptscriptstyle R}_{\lower 2pt \hbox{$\scriptstyle i$}})^{2} 
  \;.
 \label{vmqc}
\end{equation}
Note that in this case one searches for the optimal pairing of red and
blue quarks---independent of the position of the green quarks; one
then repeats the procedure to find the corresponding optimal
blue-green and green-red pairings. An exhaustive approach to the
pairing problem requires of a search among $A!$ different
configurations.  However, efficient pairing algorithms---with a
computational cost proportional to $N^{3}$---have been already
developed by mathematicians\cite{burder80} and implemented by
economists; economists have long been interested in the problem of
pairing $N$ factories with $N$ retail stores in order to minimize the
overall cost of exchanging goods. In spite of the change, both models
share many common features. Clearly, they are identical in the case of
an isolated cluster, and, thus, also in the very-low density
limit. Moreover, both models guarantee cluster separability, thus
avoiding the emergence of van der Waals forces. However, there are
some differences. Most notoriously, by pairing quarks independently
there is no guarantee that the quarks will be grouped into three-quark
clusters; color-neutral hadrons in this model may contain any
multipole of three quarks (see Fig~\ref{figthree}).
 We have computed the energy per quark and variational parameter
as a function of density for a total of $N=96$ quarks (no figure
is shown).  We have identified a discontinuity in the derivative 
of the energy and an abrupt transition in the variational 
parameter\cite{horpie92}. This transition from nuclear- to
quark-matter is accompanied by a dramatic change in the
quark pairings. To characterize this transition we define an $n$-quark
cluster probability (with $n$ any multipole of 3) as the probability
of finding a quark as part of an $n$-quark cluster.  In
Fig.~\ref{figcluster} we display the $n$-quark cluster probabilities
as a function of density. In the low-density nuclear phase most of the
quarks are clustered into simple nucleons containing three quarks
each. In contrast, the 3-quark cluster probability drops to about one
third in the quark-matter phase. The 6-, 9-, and 12-quark cluster
probabilities are small at all densities. The majority of the
probability resides in large clusters involving 15 or more quarks.
Indeed, the size of these clusters are comparable to the simulation
volume. This transition---in which the strings grow and fill the
simulation volume---is analogous to a percolation phase transition
observed in some condensed-matter systems. Clearly, our cluster
probabilities are model dependent.  Indeed, they depend sensitively on
the definition of the many-quark potential [see
Eq.~(\ref{vmqc})]. Yet, these quark-cluster probabilities were
predicted within our model; this is in contrast to the majority of
theoretical approaches which treat multi-quark cluster probabilities
as arbitrary parameters.

\section{Dynamic Properties}\label{sec:dynamic}
 In this section we report on our progress towards calculating
the exact dynamic response of hadronic matter. To our knowledge, 
this is the first time that an attempt has been made at modeling 
the response of a many-body system with confined degrees of freedom.
We are interested in computing the longitudinal (or charge) response
of the system,
\begin{equation}
   S_{L}({\bf q},\omega)=\sum_{n} 
   \Big|\langle\Psi_{n}|\hat{\rho}({\bf q})|\Psi_{0}\rangle\Big|^{2}
   \delta\Big[\omega-(E_{n}-E_{0})\Big] \;.
 \label{sqw}
\end{equation}
Here $\hat{\rho}$ is the charge-density operator, $\Psi_{0}$ is the 
exact ground state of the system, and $\Psi_{n}$ is an excited state 
with excitation energy $E_{n}-E_{0}$. Note that the response is probed
at a momentum transfer ${\bf q}$ and at an energy loss $\omega$. To 
simulate the response it is convenient to introduce its Fourier 
transform---the autocorrelation function:
\begin{equation}
  S_{L}({\bf q},t) = \int_{0}^{\infty} 
   d\omega e^{-i\omega t}S_{L}({\bf q},\omega) =
   \langle\Psi_{0}|\hat{\rho}^{\dagger}({\bf q})
   e^{-i(\hat{H}-E_{0})t}\hat{\rho}({\bf q})|\Psi_{0}\rangle \;.
 \label{sqt}
\end{equation}
This form is suitable to be simulated in the computer, but only 
in Euclidean (or imaginary) time: $t\rightarrow -i\tau$; the main
difficulty in simulating the response in real time originates 
from its oscillating phase. It is interesting to note that the 
same imaginary-time evolution operator that appears in the
calculation of the response can be used to compute the exact 
ground state of the system\cite{cnk84}
\begin{equation}
  \lim_{\tau\rightarrow\infty}e^{-(\hat{H}-E_{0})\tau}
  |\Phi\rangle = |\Psi_{0}\rangle \langle\Psi_{0}|\Phi\rangle \;. 
 \label{psigs}
\end{equation}
Thus, the Euclidean response of the system can be calculated
exactly, up to statistical uncertainties. However, two serious
challenges remain to be addressed: a) how to analytically continue 
the Euclidean response to real time and b) how to deal with the
large statistical uncertainties that are ubiquitous to the simulation
of fermionic systems. The first problem emerges from the realization
that the inverse Laplace transform of numerical data represents an 
ill-posed problem. A promising tool in the reconstruction of the
real-time response is the method of maximum entropy\cite{skill89}. 
Maximum entropy uses a Bayesian approach to the problem; it finds 
the most probable real-time response, $S_{L}({\bf q},\omega)$, that 
has the computed Euclidean response as its Laplace transform. In 
essence, maximum entropy represents a sophisticated chi-square minimization 
procedure---aided by physics (in the form of a model and sum rules) 
and assumptions about smoothness and positivity. The second problem
represents a formidable challenge, indeed. The essential difficulty 
arises from cancelling (plus and minus) signs associated with 
fermionic determinants\cite{negorm88}. Although useful strategies 
might exist, these cancellations can not be avoided in more than one 
spatial dimension. In the special case of one dimension the sign 
cancellation problem can be circumvented by working in an ordered 
subspace in which the wave function can be chosen to be positive 
definite; as a quark attempts to move pass one of its neighbors, 
the emergence of a large and repulsive ``Pauli potential'' precludes 
the move. Hence, once ordered, the quarks will remain order throughout
the simulation. Because of the formidable challenges encountered by
exact three-dimensional calculations, we concentrate for the rest of 
this contribution on the exact simulation of many-quark systems in one
spatial dimension. Moreover, we will also avoid discussing the
analytic continuation of the Euclidean response. Thus, in this 
contribution we focus exclusively on the Coulomb sum $S_{L}(q,t=0)$. 

 For the simulations we employ the original string-flip model 
developed by Lenz and collaborators\cite{lenz86}. Here, quarks 
are restricted to one spatial dimension and are devoid of any 
intrinsic degrees of freedom. Moreover, quarks are confined within 
two-quark hadrons by harmonic forces. In spite of its simplicity, 
the model shows rich and interesting behavior associated with the 
internal quark substructure of the hadron\cite{hmn85,horpie91}. 
It is the impact of this quark substructure on the response that 
we now address. 
\begin{figure}[h]
\begin{center}
 \null
 \vskip1.0in
 \includegraphics{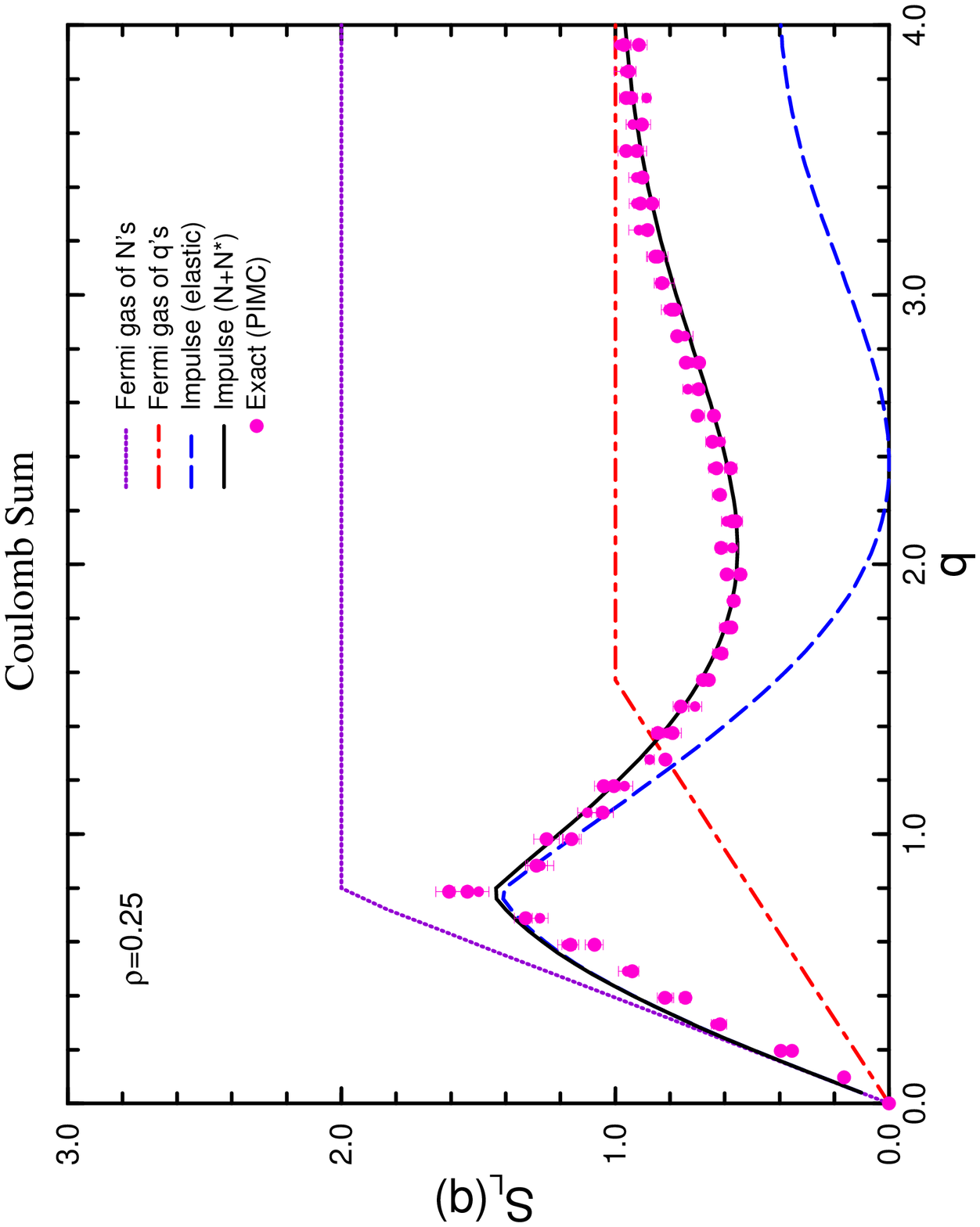}
 \vskip2.0in
\fcaption{Coulomb sum as a function of momentum transfer.}
\label{figcoulomb}
\end{center}
\end{figure}
In Fig.~\ref{figcoulomb} we display the Coulomb sum as a function of 
the momentum transfer for a density of $\rho=0.25$. For this density 
the confinement scale is considerably smaller than 
the average inter-hadron separation; thus a nucleon at this 
density---although slightly ``swollen''---resembles to a very good 
approximation a nucleon in free space. The dot-dashed line represents
the response of free Fermi gas of quarks. It is evident that for
small $q$ the Fermi-gas response is well below the exact response 
of the system. This behavior suggests that at these momentum transfers 
both quarks in a hadron respond coherently. However, the coherence is 
incomplete, as the response of the system is below that of a free
Fermi gas of nucleons (dotted line). Some of the coherence is lost 
because of the intrinsic quark substructure of the nucleon; the
nucleon ``survival'' probability is proportional to its elastic
form factor. Hadronic descriptions of the nuclear response rely on 
the impulse approximation; one assumes that the probe couples to  
a point-nucleon current multiplied by the elastic form factor
of the nucleon---which remains unchanged from its free-space
value. The long-dashed line depicts the outcome of such a
calculation. Note that here we have assumed that the point-nucleon
response is that of a free Fermi gas of nucleons. The impulse
approximation gives, indeed, an excellent description of the numerical
data for small momentum transfers. However, as the momentum transfer
becomes comparable to the inverse nucleon size, there is substantial 
strength in the exact response that the simple impulse-approximation
approach can not account for. This extra strength must be contained 
in the nucleon resonances. To test such a scenario we introduce an
extended impulse-approximation approach that incorporates all nucleon 
resonances. In this limit, the dynamic response of the system is 
proportional to the product of the Fermi-gas response of ``point'' 
nucleons times the single-nucleon response: 
\begin{equation}
  {1 \over N}S_{\scriptscriptstyle EIA}(q,\tau) = 
  {1 \over 2}e_{\scriptscriptstyle N}^{2}
  \left({S_{A}^{FG}(q,\tau) \over A}\right)
  \sum_{n=0}^{\infty} e^{-2n\omega_{0}\tau} 
  \Big|F_{n0}(q)\Big|^{2} \;.
 \label{eia}
\end{equation}
Here $F_{n0}(q)$ is a single-nucleon transition form factor, 
$\omega_{0}$ is the oscillator frequency, and 
$e_{\scriptscriptstyle N}^{2}=4$ is the square of the nucleon 
charge (the quark charge has been defined to be one). Note that 
some of the dimensionful parameters have been temporarily restored. 
The outcome of such a calculation is shown with the solid line. 
The agreement with the exact numerical calculation is excellent over 
the entire momentum-transfer range. The widely used impulse 
approximation, which is obtained from the above expression by 
retaining only the $n=0$ term, misses most of the intermediate- and 
high-$q$ strength. The extra strength is, indeed, contained in the 
excitation of the nucleon resonances.

\section{Conclusions}\label{sec:concl}
 We have examined the role of the quark substructure of hadrons on
nuclear observables using simple constituent quark models. The
string-flip models used here confine quarks within hadrons, enable the
hadrons to separate without generating van der Waals forces, and are
explicitly symmetric in all quark coordinates. The crucial feature of
these models is the need to determine an optimal grouping of quarks
into hadrons. In a three-dimensional model with a (global) color
degree of freedom we have identified a nuclear- to quark-matter
transition---characterized by a dramatic rearrangement of strings. We
have also calculated the exact Coulomb sum of a many-quark system in
one spatial dimension. At low density and small momentum transfers the
response was considerably smaller than that of a free Fermi gas of
quarks; this suggests a coherent response from all the quarks in the
hadron. However, this coherence was incomplete, as the internal quark
substructure of the hadron led to a suppression of the response
relative to that of a free Fermi gas of nucleons. As the momentum
transfer increased and became comparable to the inverse nucleon size,
substantial strength above the impulse-approximation limit was clearly
identified. We have concluded that most of this extra strength was
contained in the excitation of the nucleon resonances.  

 In the future, we would like to include additional intrinsic
(spin-isospin) degrees of freedom into our simulations. We will 
also continue our search, via the maximum-entropy method, for 
quark giant resonances. Further, we could easily add strange quarks 
into our simulations by assuming a flavor-independent form for the 
potential; one obtains the optimal grouping of quarks into hadrons
irrespective of the flavor of the quarks. This will enable us to 
study the transition to strange matter. In summary, we have shown 
that, in spite of their apparent simplicity, string-flip models 
of nuclear matter display rich behavior that could result in valuable 
insights into the role of nucleon substructure in hadronic physics.

\section{Acknowledgments}\label{sec:acknol}
The work reported in this contribution was done in collaboration
with G.M. Frichter, S. Gardner, C.J. Horowitz, and W. Melendez.
This work was supported by the U.S. Department of Energy under
Contracts Nos. DE-FC05-85ER250000 and DE-FG05-92ER40750.


\section{References }\label{sec:refs}
\vspace{-0.7cm}
\begin{thebibliography}{99}
\bibitem{lenz86}   F. Lenz, J.T. Londergan, E.J. Moniz, 
                   R. Rosenfelder, M. Stingl, and K. Yazaki, 
                   {\it Ann. Phys.} {\bf 170} (1986) 65.
\bibitem{isgkar79} N. Isgur and G. Karl, 
		   {\it Phys. Rev.} {\bf D20} (1979) 1191.
\bibitem{grelip81} O.W. Greenberg and H.J. Lipkin, 
                   {\it Nucl. Phys.} {\bf A370} (1981) 349.
\bibitem{hmn85}    C.J. Horowitz, E. J. Moniz, and J.W. Negele,
                   {\it Phys. Rev.} {\bf D31} (1985) 1689.
\bibitem{watson89} P.J.S. Watson, 
                   {\it Nucl. Phys.} {\bf A494} (1989) 543;
                   A.B. Migdal and P.J.S. Watson, 
                   {\it Phys. Lett.} {\bf 252B} (1990) 32. 
\bibitem{horpie91} C.J. Horowitz and J. Piekarewicz, 
                   {\it Phys. Rev.} {\bf C44} (1991) 2753.
\bibitem{horpie92} C.J. Horowitz and J. Piekarewicz, 
                   {\it Nucl. Phys.} {\bf A536} (1992) 669.
\bibitem{fripie94} G.M. Frichter and J. Piekarewicz,
                   {\it Comput. Phys.} {\bf 8} (1994) 223.
\bibitem{alber92}  W.M. Alberico, M.B. Barbaro, A. Molinari,
                   and F. Palumbo, 
                   {\it Z. Phys.} {\bf A341} (1992) 327;
                   W.M. Alberico, M.B. Barbaro, A. Magni, 
 		   and M. Nardi,
                   {\it Nucl. Phys.} {\bf A552} (1993) 495.
\bibitem{burder80} R.E. Burkard and U. Derigs, {\it Lecture
                   Notes in Economics and Mathematical Systems} 
                   (Springer-Verlag, Berlin, 1980), vol. 184. 
\bibitem{cnk84}    S.A. Chin, J.W. Negele, and S.E. Koonin,
                   {\it Ann. Phys.} {\bf 157} (1984) 140.
\bibitem{skill89}  J. Skilling, in {\it Maximum Entropy and
		   Bayesian Methods} (Klumer Academic Publisher, 1989)
		   p. 45.
\bibitem{negorm88} J.W. Negele and H. Orland, 
		   {\it Quantum Many-Particle Systems}
		   (Addison-Wesley, Redwood City, 1988).
\end{thebibliography}
\end{document}